\documentclass{aa}

\usepackage{graphicx}
\usepackage{amssymb}
\usepackage{latexsym}
\usepackage{latexsym}
\usepackage{graphics}
\usepackage{txfonts}
\usepackage{natbib}

\def\ha{H$\alpha$}
\def\hb{H$\beta$}

\def\niib{[NII]$\lambda$6584}
\def\niia{[NII]$\lambda$6548}

\def\oiiia{[OIII]$\lambda$4959}
\def\oiiib{[OIII]$\lambda$5007}

\def\siia{[SII]$\lambda$6716}
\def\siib{[SII]$\lambda$6730}
\def\siiab{[SII]$\lambda$6716,30}
\def\oi{[OI]$\lambda$6300}

\begin{document}


\title{The MAGNUM survey: Positive feedback in the nuclear region of NGC~5643 suggested by MUSE
		\thanks{This work is based on observations made at the European Southern Observatory, Paranal, Chile (ESO program 60.A-9339)}
}

\author{G. Cresci\inst{1}\fnmsep, 
A. Marconi\inst{2},
S. Zibetti\inst{1},
G. Risaliti\inst{1},
S. Carniani\inst{2},
F. Mannucci\inst{1},
A. Gallazzi\inst{1},
R. Maiolino\inst{3,} \inst{4},
B. Balmaverde\inst{2},
M. Brusa\inst{5,}\inst{6},
A. Capetti\inst{7},
C. Cicone\inst{8},
C. Feruglio\inst{9,}\inst{10,}\inst{11},
J. Bland-Hawthorn\inst{12},
T. Nagao\inst{13},
E. Oliva\inst{1},
M. Salvato\inst{14},
E. Sani\inst{15},
P. Tozzi\inst{1},
T. Urrutia\inst{16},
G. Venturi\inst{2}
}

\institute{INAF - Osservatorio Astrofisco di Arcetri, largo E. Fermi 5, 50127 Firenze, Italy \\ \email{gcresci@arcetri.astro.it}  
	\and
	Universit\`a degli Studi di Firenze, Dipartimento di Fisica e Astronomia, via G. Sansone 1, 50019 Sesto F.no, Firenze, Italy
	\and
	Cavendish Laboratory, University of Cambridge, 19 J.J. Thomson Ave., Cambridge, UK 
	\and
	Kavli Institute for Cosmology, University of Cambridge, Madingley Road, Cambridge, UK 
	\and
	Dipartimento di Fisica e Astronomia, Universit\'a di Bologna, viale Berti-Pichat 6/2, 40127 Bologna, Italy
	\and
	INAF - Osservatorio Astronomico di Bologna, via Ranzani 1, 40127 Bologna, Italy 
	\and
	INAF-Osservatorio Astrofisico di Torino, via Osservatorio 20, 10025 Pino Torinese, Italy
	\and
	ETH Z\"urich, Institute for Astronomy, Department of Physics, Wolfgang-Pauli-Strasse 27, 8093 Zürich, Switzerland
	\and
	Scuola Normale Superiore, Piazza dei Cavalieri 7, I-56126 Pisa, Italy
	\and
	IRAM - Institut de RadioAstronomie Millim\'etrique, 300 rue de la Piscine, 38406 Saint Martin d’H\`eres, France
	\and
	INAF - Osservatorio Astronomico di Roma, via Frascati 33, 00044 Monte Porzio Catone (RM) Italy
	\and
	Sydney Institute for Astronomy, School of Physics, University of Sydney, NSW 2006, Australia
	\and
	Research Center for Space and Cosmic Evolution (RCSCE), Ehime University, Bunkyo-cho 2-5, Matsuyama, Ehime 790-8577, Japan
	\and
	Max Planck Institut f\"ur extraterrestrische Physik, Postfach 1312, 85741 Garching, Germany
	\and
	ESO, Alonso de Cordova 3107, Casilla 19, Santiago 19001, Chile
	\and
	Leibniz Institut f\"ur Astrophysik, An der Sternwarte 16, 14482 Potsdam, Germany\\
}

\date{Received ; accepted }

\abstract{We study the ionization and kinematics of the ionized gas in the nuclear region of the barred Seyfert 2 galaxy NGC~5643 using MUSE integral field observations in the framework of the MAGNUM (Measuring Active Galactic Nuclei Under MUSE Microscope) survey. The data were used to identify regions with different ionization conditions and to map the gas density and the dust extinction. We find evidence for a double sided ionization cone, possibly collimated by a dusty structure surrounding the nucleus. At the center of the ionization cone, outflowing ionized gas is revealed as a blueshifted, asymmetric wing of the [OIII] emission line, up to projected velocity $v_{10}\sim-450\ km\ s^{-1}$. The outflow is also seen as a diffuse, low luminosity radio and X-ray jet, with similar extension.
The outflowing material points in the direction of two clumps characterized by prominent line emission with spectra typical of HII regions, located at the edge of the dust lane of the bar. We propose that the star formation in the clumps is due to `positive feedback' induced by gas compression by the nuclear outflow, providing the first candidate for outflow induced star formation in a Seyfert-like radio quiet AGN. This suggests that positive feedback may be a relevant mechanism in shaping the black hole-host galaxy coevolution.}

\keywords{Galaxies: active -- Galaxies: individual (\object{NGC5643}) - ISM: jets and outflows -- Techniques: imaging spectroscopy}

\authorrunning{Cresci et al.}
\titlerunning{Positive feedback in the nuclear region of NGC~5643 suggested by MUSE}

\maketitle

\section{Introduction}
Active Galactic Nuclei (AGN) have a profound influence on their host galaxies, as they are capable of ionising large fractions of the interstellar medium (ISM) and of accelerating fast outflows. These are powerful enough both to sweep away most of the gas in the ISM and to heat the gas in the surrounding halo, inhibiting gas accretion (``negative feedback'', see Fabian et al. \citealp{fabian12} and references therein). Fast outflows could also induce star formation (``positive feedback'') in the pressure compressed molecular clouds (e.g. Silk et al. \citealp{silk13}) or in the outflowing gas (Ishibashi \& Fabian \citealp{ishibashi12},\citealp{ishibashi14}, Zubovas \& King \citealp{zubovas14}).
\begin{figure*}
	\begin{center}
		\includegraphics[width=1\textwidth]{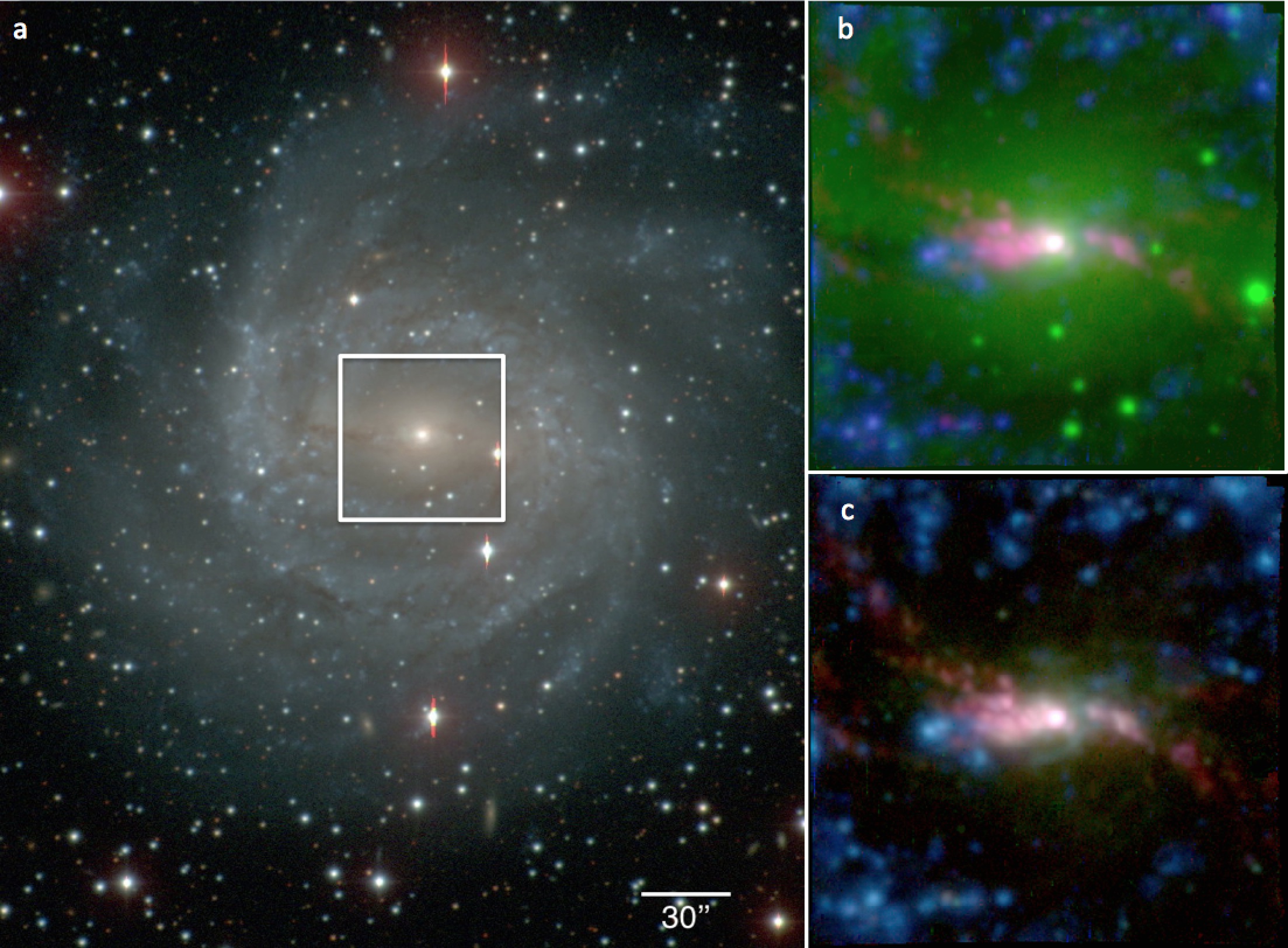}
	\end{center}
	\caption{The galaxy NGC~5643. \textit{Panel a:} Three color (Blue: B band; Green: V band; Red: I band) image of the galaxy obtained with the 2.5m du Pont Telescope for the Carnegie-Irvine Galaxy Survey (CGS, Ho et al. \citealp{ho11}). The MUSE field of view is shown as a white square. North is up and East is left. \textit{Panel b:} Three colors MUSE map, showing \ha\ emission in Blue, \oiiib\ emission in Red and the integrated continuum emission from 5200 to 6000 $\AA$ in Green. The tip of the dust lane in the bar, visible on larger scale in Panel a, is evident next to the blue clump East of the nucleus as a deep in the continuum. \textit{Panel c:} Three colors MUSE maps, showing \ha\ in Blue and \oiiib\ in Red as before, and \niib\ emission in Green. The two-sided AGN ionization cone traced by the filamentary [OIII] emission is prominent, as well as the strongly \ha\ emitting clumps between the two eastern lobes of the cone. Nord is up and East is left.} 
	\label{3colors}
\end{figure*}

Although such `positive feedback' has been invoked in recent years by several theoretical works to explain observed correlations between, e.g., AGN luminosities and nuclear star formation rates (Imanishi et al. \citealp{imanishi11}, Zinn et al. \citealp{zinn13}, Zubovas et al. \citealp{zubovas13}), black hole accretion and star formation rates in AGN (Silverman et al. \citealp{silverman09}, Mullaney et al. \citealp{mullaney12}, Silk \citealp{silk13}) or downsizing of both black holes and spheroids (Silk \& Norman \citealp{silk09}), still very few observations of such phenomenon at work have been argued for, mostly related with spatial alignment of star forming regions or companionion galaxies with very powerful radio jets (e.g. Kramer et al. \citealp{klamer04}, Croft et al. \citealp{croft06}, Elbaz et al. \citealp{elbaz09}, Feain et al. \citealp{feain07}, Crockett et al. \citealp{crockett12}, Salom\'e, Salom\'e \& Combes \citealp{salome15}). The only known example of ``positive'' feedback at high-z in a radio quiet AGN has been recently presented by Cresci et al. \citep{cresci15}, who detected with near-IR Integral Field observations star forming clumps in the host galaxy of a $z\sim1.6$ quasar, possibly triggered by the outflow pressure at its edges. However, we are still lacking observations of positive feedback at lower AGN luminosity and outflow energy, in order to understand its importance in normal AGNs. 

In this paper we present observations of the nuclear region of NGC~5643, a barred, radio-quiet Seyfert 2 galaxy at a distance of $\sim17.3$ Mpc seen almost face on ($i\sim-27 \pm 5\deg$, de Vaucouleurs et al. \citealp{RC2cat76}), with a well known ionization cone extending eastward of the nucleus, parallel to the bar (Schmitt et al. \citealp{schmitt94}, Simpson et al. \citealp{simpson97}, Fischer et al. \citealp{fischer13}).  A three color image (B,V, I bands) from the Carnegie-Irvine Galaxy Survey (CGS, Ho et al. \citealp{ho11}) is shown in Fig.~\ref{3colors}, panel \textit{a}. As usually observed in barred spiral galaxies, sharp, straight dust lanes extend from the sides of the central nucleus out to the end of the bar, roughly parallel to its major axis and being more clearly visible to the East of the nucleus. The galaxy is also known to host a diffuse, low luminosity radio jet on both sides of the nucleus, which is nearly $30"$ long ($2.5\ kpc$, Leipski et al. \citealp{leipski06}, Morris et al. \citealp{morris85}). 

This galaxy was observed with the Multi Unit Spectroscopic Explorer MUSE, the optical large field integral-field spectrometer at the VLT (Bacon et al. \citealp{bacon10}) as part of the science verification run, in the framework of the MAGNUM (\textit{Measuring Active Galactic Nuclei Under MUSE Microscope}) survey, a program aimed at the observation of nearby Active Galactic Nuclei (AGNs), to study the physical conditions of the Narrow Line Regions (NLRs), the interplay between nuclear activity and star formation, and the effects and acceleration mechanisms of outflows.

In the following we adopt the systemic velocity measured by Koribalski et al. \citep{koribalski04}, $1199\ km\ s^{-1}$. We use $H_0=69.6\ km\ s^{-1}\ Mpc^{-1}$ and $\Omega_M=0.286$ (Hinshaw et al. \citealp{hinshaw13}). At the distance of the galaxy of $17.3\ Mpc$, the angular scale is $1"=83\ pc$.

\section{Observations, data reduction and analysis}

NGC~5643 was observed with MUSE during the science verification run on June 24, 2014, under program 60.A-9339 (PI Marconi/Hawthorn/Salvato). The average seeing during the observations, derived directly from foreground stars in the final datacube, was $FWHM=0.88"\pm0.02"$. The nuclear region of the galaxy was observed with four dithered pointings of $500\ s$ each, with the sky sampled with 4 shorter $100\ s$ exposures in between.
The data reduction was performed using the recipes from the early-release MUSE pipeline (version 0.18.1), as well as a collection of custom IDL codes developed to improve the sky subtraction, response curve and flux calibration of the data. Further details on the data reduction will be provided in a forthcoming paper (Carniani et al., in preparation). 
The final datacube consists of $315\times315$ spaxels, for a total of almost 100000 spectra with a spatial sampling of $0.2"\times0.2"$ and a spectral resolution going from 1750 at $465\ nm$ to 3750 at $930\ nm$.
\begin{figure*}
	\begin{center}
		\includegraphics[width=0.8\textwidth]{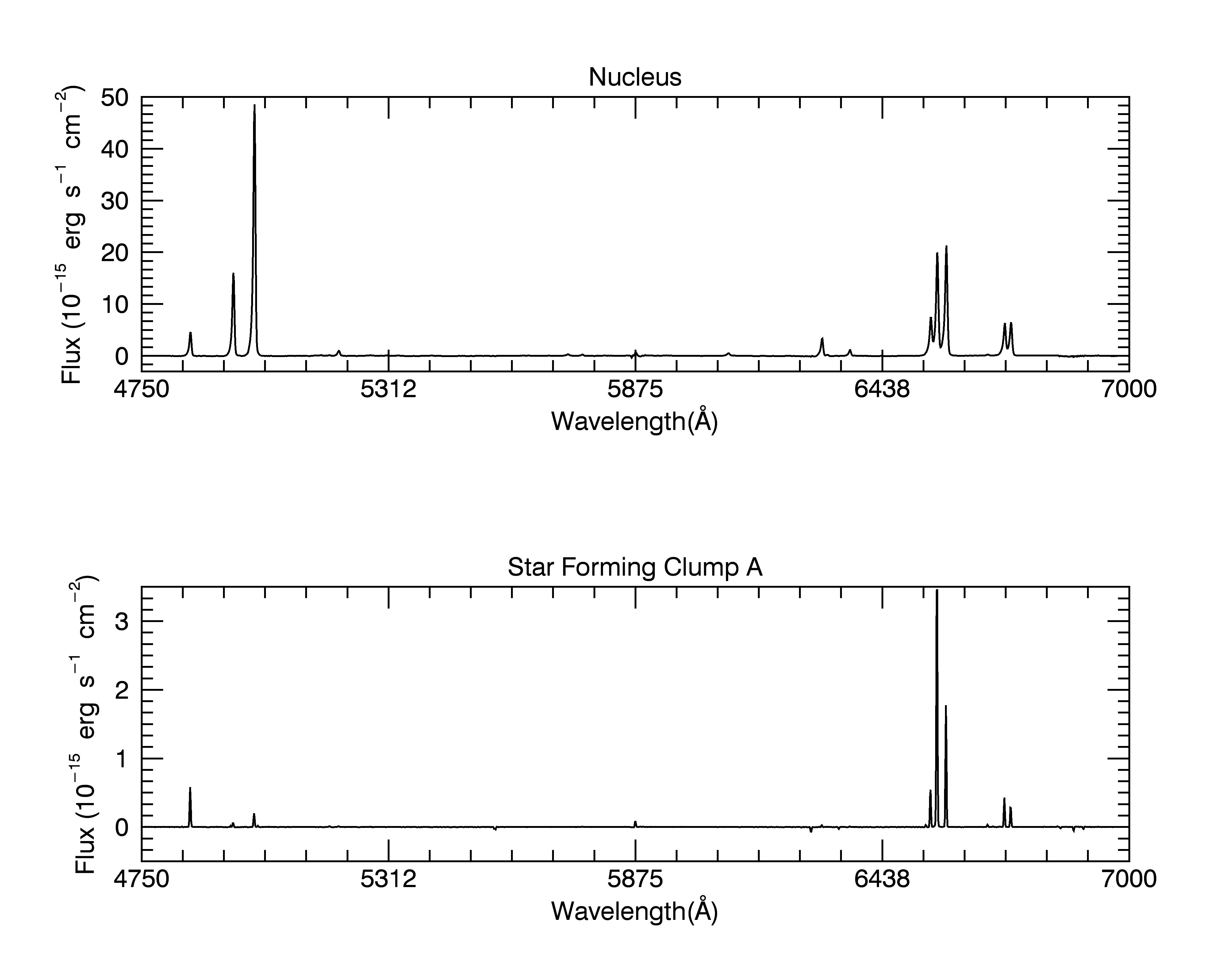}
	\end{center}
	\caption{Continuum subtracted, rest frame spectra extracted from a region of $1.6"\times1.6"$ around the nucleus (\textit{upper panel}) and around the star forming clump A (\textit{lower panel}). The different excitation between the two regions is evident from the different line ratios, where [OIII] and [NII] are brighter in the nucleus, while stronger \ha\ and \hb\ emission is detected in the star forming clumps.
	}
	\label{spectra}
\end{figure*}

Moreover, NGC~5643 was observed in the X-rays by the Chandra Observatory on December 26, 2004, for 7.5 ks with the Advanced CCD Imaging Spectrometer (ACIS: Garmire et al. \citealp{garmire03}). Data were reduced with the Chandra Interactive Analysis of Observations 4.5 (CIAO; Fruscione et al. \citealp{fruscione06}) and the Chandra Calibration Data Base 4.6.5, adopting standard procedures. In order to increase the spatial resolution, the imaging analysis was performed by applying a factor of 2 sub-pixel event repositioning 
(e.g. Wang et al. \citealp{wang11}). We therefore used a pixel size of $0.246"$, instead of the native $0.495"$. 

\subsection{Modeling the stellar continuum and emission lines}

In presence of young/intermediate-age stellar populations, which
present intense Balmer absorption (e.g. $\mathrm{EW(H}\beta)$ can
reach up to 8\AA), a careful subtraction of the underlying stellar
continuum is important in order to recover the correct emission line fluxes and profiles. 
Although the optimal approach is to make a simultaneous fit
of the stellar continuum and of the emission lines (see e.g. Sarzi et al. \citealp{sarzi06}), 
with $10^5$ spaxels and a spectral sampling of $1.25\AA$
this becomes computationally unfeasible for a MUSE datacube on a
standard desktop computer. On the other hand, considering the
intensity of the emission lines we aim at studying here (see next Section), sufficient accuracy can be obtained by
fitting the stellar continuum while avoiding the regions interested by
emission lines and then by extrapolating the fit over those regions.
We use the pPXF code (Cappellari \& Emsellem \citealp{cappellari04}) to perform this task over the
wavelength range of interest, i.e. shortward of $7000\AA$. A linear
non-negative combination of stellar spectral templates is fitted to
the data by adjusting the systemic velocity and the velocity
dispersion together with the coefficients of the linear combination.
We consider 40 stellar templates from the simple stellar populations provided by Bruzual \& Charlot \cite{bc03}), distributed in a grid of 10 ages (5.2, 25, 100, 290, 640, 900 Myr, 1.4, 2.5, 5.0 and 11 Gyr) and 4 metallicities (0.2, 0.4, 1, and 2.5 times solar). Each spectral template is converted from wavelength to velocity space and so is each observed spectrum. pPXF iteratively adjusts the systemic velocity and the gaussian velocity dispersion of each template (this is done by plain convolution in the velocity space with a gaussian kernel; velocity and dispersion are assumed to be common to all component templates) and computes the coefficients of the linear combination of templates that best reproduce the observed spectrum. The procedure is iterated until the minimum $\chi^2$ and the best fitting systemic velocity, velocity dispersion and coefficients of the template linear combination are obtained (see Cappellari \& Emsellem \citealp{cappellari04} for more details). Note that the template distribution in age and metallicity allows to cover virtually any (combination of) stellar populations inside a galaxy, including the very young ages that characterise the star-froming blobs. 
We mask regions corresponding to $\pm 750
\mathrm{km~s}^{-1}$ around the main astrophysical emission lines and
around the Na\textsc{I} absorption at rest and in the galaxy's
restframe to avoid contamination from the interstellar absorption.
We note that the fit is expected to be definitely degenerate in terms of the possible linear combinations of SSPs, but this is irrelevant to our goal: a realistic model of the star formation history and metallicity distribution is not required to ensure just an accurate stellar continuum subtraction. We verify that we obtain an accurate subtrction by comparing the residuals from the fit with the estimated noise level and finding them consistent with each other. 

This fitting procedure produces reliable results only if a sufficient
signal-to-noise ratio (SNR) in the continuum is provided, namely $>10$
per \AA. In the original datacube at full
resolution this is achieved only in the brightest parts; for the fainter regions we
apply an adaptive smoothing procedure which extends the
\textsc{adaptsmooth} code (Zibetti \citealp{zibetti09a}, Zibetti et al. \citealp{zibetti09b}) to work with
spectral datacubes (\textsc{azmooth3}, Zibetti et al. in preparation).
At each spaxel the SNR is estimated in a narrow wavelength window on a
featureless continuum, between 6000 and 6050 \AA. If the SNR is lower
than the threshold of 20 per spectral pixel, concentric annuli of
spaxels are coadded iteratively until the SNR of 20 is reached. No
more than 10 annuli are coadded at each position. The smoothed
cube warrants sufficient SNR at all spaxels while preserving as much
as possible the effective spatial resolution. 
The typical smoothing radius (in pixels, 1 pix=$0.2"$) varies from 1 (i.e. no smoothing) within  $5"$ from the centre to 2, 3, and 5 at distances of $\sim13"$, $18"$, and $25"$, respectively. Larger smoothing radii are required only in sparse regions, typically outside $25"$. Indeed, most of the analysis presented in this work is unaffected or marginally affected by the smoothing: the EELR are in fact mostly contained in the no-smoothing zone or in the 2-pixel radius smoothing zone. In any case, the smoothing affects scales typically $<2"$ in diameter, hence smaller than those we investigate in the following.
The output is then processed with pPXF as explained above to produce a pure stellar
continuum cube, stellar velocity and stellar velocity dispersion maps.

\subsection{Emission line fitting, density and extinction maps} \label{lines}

The obtained stellar continuum is subtracted from the original data, and the resulting datacube is used for the following analysis. We fit the main emission lines accessible in the MUSE wavelength range, i.e. H$\beta$, \oiiia, \oiiib, \oi, \niia, \ha, \niib, \siia, \siib. To better reproduce the observed spectral profiles in the data, we use both a two Gaussian components fit and a single Gaussian fit to each line. The flux obtained with the double Gaussian is used in the following only for the spaxels where the second component has a $SNR>3$ in the fit, i.e. in the nuclear region where the line profile is more complex (see Fig.~\ref{spectra}). This choice allows us to use the more degenerate two components fit only where it is really needed. 
\begin{figure}
	\begin{center}
		\includegraphics[width=0.5\textwidth]{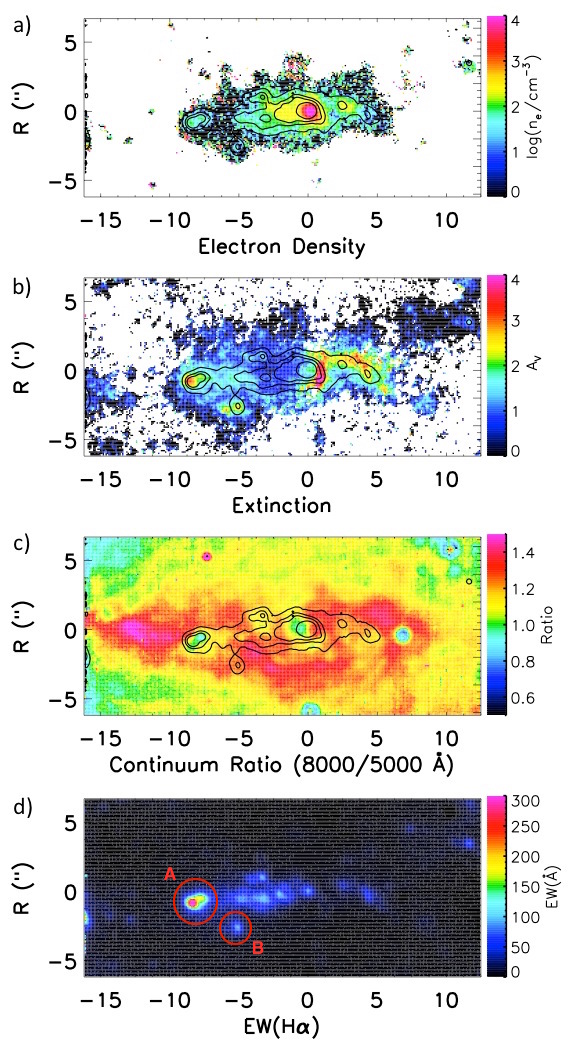}
		\caption{Electron density (\textit{panel a}) and extinction map (\textit{panel b}), as derived from the \siia/\siib\ and \ha/\hb\ line ratios (see text) with \ha\ contours overplotted. The star forming clump has higher density and extinction than its surroundings. A dusty structure with very high extinction is detected just West of the nucleus and H$\alpha$ peak, corresponding to the tip of the bar driven dust lane. 
		Both maps are displaying the spaxels with $SNR>5$. 
		The ratio between the continuum flux in the $8000-8500\ \AA$ spectral region and the $5100-5500\ \AA$ one, with \ha\ contours overplotted, is shown in \textit{panel c}. While the star forming clumps and the nucleus are bluer due to the intrinsic shape of the continuum, the higher extinction regions are showing redder colors, tracing the dust distribution in the regions even where the SNR of the emission lines is lower. The dust lanes to the W and E of the nucleus, an extinction ridge to the S connecting the dust lane with the dust structure surrounding the nucleus are clearly visible.
		\textit{Panel d} shows the rest frame H$\alpha$ Equivalent Width map. The star forming clump A, marked with a circle as well as clump B, shows the highest $EW(\textrm{H}\alpha)=350\ \AA$ in the field of view, suggesting an age of $\sim10^7$ yrs (see text for details).}
	\label{neav}
	\end{center}
\end{figure}

From these fluxes we constructed emission line maps in the \ha\, [OIII] and [NII] transitions (see Fig.~\ref{3colors}b,c). These maps are already clearly showing that the line emitting gas presents an elongated morphology, aligned with the bar in the E-W direction. We will refer in the following to this region as Extended Emission Line Region (EELR). As reported by Schmitt et al. \cite{schmitt94}, the EELR is elongated both to the E and to the W of the nucleus, although it is much fainter to the W, especially at shorter wavelengths (e.g. Fig.~\ref{3colors}, panels, b and c). As we will discuss later, this is due to the higher extinction to the W, indicating that the EELR is tilted with respect to the galaxy plane, and it is lying behind the disk of the galaxy to the W, while it is seen mostly above the disk to the E. 

An [OIII] emitting 
structure is very prominent, shown in red in Fig.~\ref{3colors}, originating at the AGN location in the galaxy center and extending in the E-W projected directions, part of an ionization bi-cone (see Fig.~\ref{3colors}, panels b and c; see also Fischer et al. \citealp{fischer13}). A V-shaped ionization cone in the core of NGC~5643 was already reported by Simpson et al. \cite{simpson97} using HST narrowband imaging, although the western part of the bi-cone was not detected in their data, probably due to lower sensitivity. 

Strikingly, a blue, \ha\ bright region is located exactly at the tip of the Eastern part of the EELR at a distance of $\sim15"=1.2\ kpc$ from the nucleus, at the location where the line emitting region reaches the dust lane of the bar (clump A, see panel b, where the dust in the bar is visible as a dip in the green continuum), and another \ha\ bright clump is located $\sim8"$ to the south-west of clump A (clump B, see also Fig.~\ref{neav} and Fig.~\ref{schema}). A comparison between the line ratios measured on clump A and the line ratio in the nuclear region shows that the excitation mechanisms are different, with the nucleus showing an AGN-like ionization as expected, while clump A displays the typical spectrum of a star forming region (see Fig.~\ref{spectra}). We will discuss further these star forming regions in the following.

We estimate the electron density from the ratio between the \siia\ and \siib\ lines, which is a widely used indicator of this quantity (e.g. Osterbrock \& Ferland \citealp{osterbrock06}). We compute the line ratio in each spaxel, and convert it to an electron density using the \textrm{IRAF} task \textrm{temden}, assuming a temperature of $10^4\ K$. The resulting electron density map is shown in Fig.~\ref{neav},a, for the regions where the [SII] lines are detected with $SNR>5$. The density is higher in the nucleus, and it decreases outwards by almost three order of magnitudes to a level of $\sim 10\ cm^{-3}$. The density is again higher ($\sim100\ cm^{-3}$) in the blue, \ha\ bright clump. We note that the density derived with this method on the nucleus is higher than the critical density for [SII] ($\sim 3 \times 10^{3}\ cm^{-3}$) above which the lines become collisionally de-excited (see Osterbrock \& Ferland \citealp{osterbrock06}). Therefore, the density derived in the nucleus with this method is very uncertain.

We also use the Balmer decrement \ha/\hb\ to derive the dust extinction map, assuming a Calzetti et al. \cite{calzetti00} attenuation law and a fixed temperature of $10^4\ K$. Although there could be in principle a large temperature range across the field of view, the expected variation in $A_V$ is $\pm0.2$ for a reasonable temperature range between 5000 K to 20000 K, typical of the conditions in HII regions and in the Narrow and Broad Line Regions of AGNs (Osterbrock \& Ferland \citealp{osterbrock06}). This do not change our qualitative picture as the difference between the dusty structure W of the nucleus and the EELR to the E is one order of magnitude greater.
the expected variation for a temperature of 5000 K or 20000 K is $\pm0.2$ in $A_V$, which do not change our qualitative picture as the difference between the dusty structure W of the nucleus and the EELR to the E is one order of magnitude greater. The resulting map is shown in Fig.~\ref{neav},b. We find that the reddening is generally higher to the W of the nucleus. This confirms that the bright line emitting gas is probably equally extended to the W and to the E of the nucleus, but while the E part is seen in front of the bar, the W side is hidden behind the bar material and suffers higher extinction. 

Moreover, the dust extinction is highest in a curved structure just W of the nucleus, which seems to be connected to the dust lane in the bar through an higher extinction ridge, visible at the bottom part of the map and around clump B (see also Fig.~\ref{neav},c, and Fig.~\ref{3colors} panel b, where the dust lane is visible as a dip in the continuum shown in green). This dusty structure appears to be $\sim7"\times1.5"$  in size, corresponding to $580\times125\ pc$, and shows the highest reddening in the central region of the galaxy, with $A_V>3$ and a maximum of $A_V=4.8$. The structure is also evident in Fig.~\ref{3colors}, as a dip in the shorter wavelength emission line map, \oiiib\ shown in red, as well in Fig.~\ref{neav}c, where the ratio between the continuum flux in the $8000-8500\ \AA$ spectral region and the $5100-5500\ \AA$ one is shown. While the star forming clumps and the nucleus are bluer due to the intrinsic shape of the continuum, the regions with higher extinction have redder colors, tracing the dust distribution in the regions even where the SNR of the emission lines is lower. In this continuum ratio map the dust lanes to the W and to the E of the nucleus are clearly visible, as well as the higher extinction ridge to the S connecting the dust lane with the dust structure surrounding the nucleus. 
It has been suggested that this dusty region represents the tip of the bar in the nuclear region, where the flowing material forms a disky structure elongated perpendicularly to the bar itself (e.g.  Morris et al. \citealp{morris85}). This is actually predicted by dynamical models, where the gas is expected to settle into a disk with rotation axis along the bar (see, e.g., Durisen et al. \citealp{durisen83}). It is possibly also responsible for the collimation of the ionizing photons along the bar (Menezes et al. \citealp{menezes15}), generating the ionization bi-cone seen in [OIII] line emission. The alignment between the inner bar and the EELR is therefore actually expected, with the ionizing continuum that escapes in this case in a cone slightly tilted with respect to the bar so that the gas is mostly illuminated in front of the bar to the E and behind the bar to the W. This high extinction region is also dominated by shock ionization, as discussed in Sect.~\ref{bptdiag}. 
\begin{figure*}[!ht]
	\begin{center}
		\includegraphics[width=0.7\textwidth]{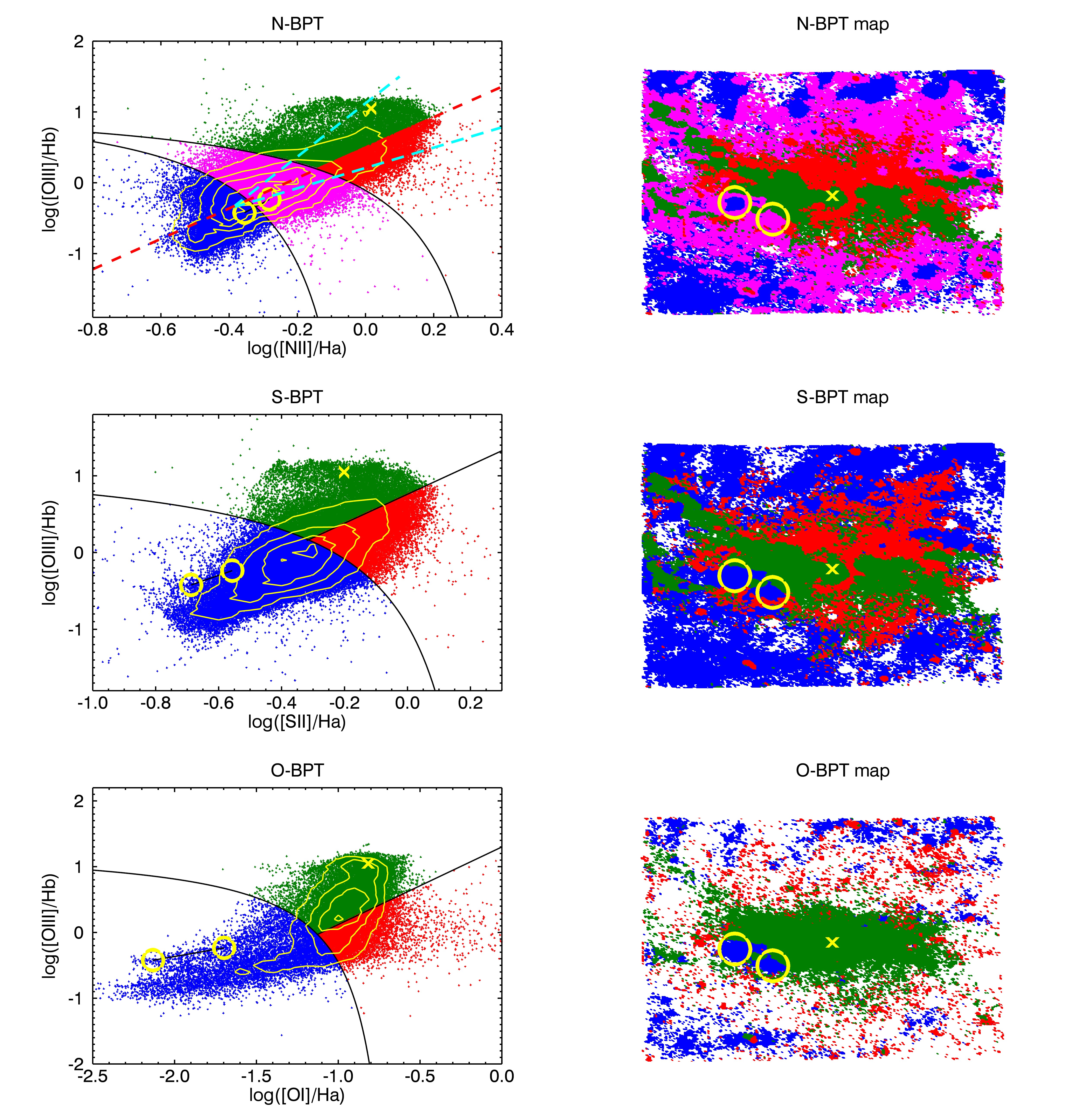}
	\end{center}
	\caption{Resolved BPT diagrams for the nuclear region of NGC~5643. The N-BPT (\niib/\ha\ vs \oiiib/\hb, \textit{upper panels}), S-BPT (\siiab/\ha\ vs \oiiib/\hb, \textit{central panels}) and O-BPT \oi/\ha\ vs \oiiib/\hb, \textit{lower panels}) diagrams for each spaxel with $SNR>3$ in each line are shown on the left, along with contours showing the density of the points on the diagram. The location of the \ha\ bright clumps A and B, as derived from an integrated spectrum extracted on a region of $1.6\times1.6"$, is shown by an yellow circle, while a cross is marking the location of the nucleus. The dashed, cyan lines in the N-BPT mark the location of points for the shock excited region of NGC~1482 and for the AGN excited emission of NGC~1365 from Sharp \& Bland-Hawthorn \cite{sharp10}. The bisector between the two, marked with a dashed red line, divide AGN dominated by shock dominated regions. A map marking each spaxel with the color corresponding to the dominant excitation at its location is shown on the right, again for spaxels with $SNR>3$. The star forming regions are marked in blue, the Seyfert-type ionization is shown in green, LINER/shock dominated regions are shown in red and intermediate regions are shown in magenta in the N-BPT. Star forming regions are prevalent in the outer part of the MUSE field of view, AGN ionization is dominating in the core and in the ionization cone, while shock excitated gas is prevalent around the EELR and the cone. The nucleus is marked with a cross, and the \ha\ bright clups A and B are marked with a yellow circle. Clump A is dominated by star formation in all the three diagrams, while Clump B is classified as intermediate by the N-BPT.}
	\label{bpt}
\end{figure*}

\section{Resolved BPT diagrams} \label{bptdiag}

We use the measured emission line fluxes in each spaxel to compute excitation maps of the galaxy region within the MUSE field of view. We consider both the classical Baldwin et al. \cite{baldwin81} BPT diagram using the \oiiib/\hb\ versus \niib/\ha\ line ratios (N-BPT in the following), as well as the alternative versions using \siiab\ (S-BPT) or \oi\ instead of \niib\ (O-BPT, see e.g. Kewley et al. \citealp{kewley06}; Lamareille \citealp{lamareille10}). In these diagrams, galaxies dominated by AGN (Seyfert-type), Low Ionization Nuclear Emission line Regions (LINERs), shocks and star formation ionization  populate different regions. For the N-BPT diagram, we also plot the location of points for two prototypical galaxies, the shock excited region of NGC~1482 and for the AGN excited emission of NGC~1365, from Sharp \& Bland-Hawthorn \cite{sharp10}. The bisector between these two fiducial traces is used to discriminate between AGN excited and shock excited gas in the N-BPT diagram.
In the right panels of Fig.~\ref{bpt} we show the maps where different regions have been colour coded according to their line excitation as derived from the corresponding BPT diagram on the left. HII regions are marked in blue, AGN dominated regions are in green, LINER-like/shock excited regions are marked in red, while composite regions in the N-BPT are shown in magenta (see Kewley et al. \citealp{kewley06}; Kauffmann et al. \citealp{kauffmann03}). 
In all BPT maps the nuclear region, the EELR and the ionization cone are dominated by AGN ionization. Shock excited gas is prevalent in the surrounding region and at the location of the dusty structure surrounding the nucleus, while star formation is dominant at the edges of the MUSE field of view, in a circumnuclear star forming ring. The only notable exception to this schematic picture are the blue, \ha\ bright regions already prominent in Fig.~\ref{3colors}. Although they are located at the tip of the central AGN dominated region and along the ionization cone, they are classified as star formation dominated by all the three BPT diagrams (clump A), or by two out of three (clump B, see Fig.~\ref{bpt} and Fig.~\ref{schema}). Interestingly, comparing the location of clump A and B in Fig.~\ref{3colors}, they appear to be also located on the ridge of the dust lane in the bar. 

Clump A is also the region with the highest \ha\ EW in teh field of view, $EW(\textrm{H}\alpha)=350\ \AA$ rest frame (see Fig.~\ref{neav},d).The other star forming regions, mostly located in a star forming ring near the edge of the MUSE field of view do not exceed $EW(\textrm{H}\alpha)=250\ \AA$. The high EW in Clump A suggests a young age of its stellar population. Using the Starburst99 models (Leitherer et al. \citealp{leitherer99}), assuming a solar metallicity and a Salpeter IMF, we obtain an estimate of the age of the cluster of $t_{cluster}=5.5 \times 10^6\ yr$ for an instantaneous burst, and $t_{cluster}=8.7 \times 10^7\ yr$ for a continuous burst of star formation. These values are to be considered as upper limits, as the measured continuum under the line is integrated over a line of sight crossing of the whole galaxy, not only of the star forming region, and therefore the measured EW(\ha) is to be considered as a lower limit to the true value.
From the measured \ha\ flux extracted in a region of $1.6"\times1.6"$ around the clump, assuming the star formation rate calibration of Kennicutt \& Evans \cite{kennicutt12}, and correcting for dust extinction using the Balmer decrement \ha/\hb\ assuming a Calzetti \cite{calzetti00} extinction law, we derive a star formation rate in the clump of $SFR_{(clump)}=0.03\ M_{\odot}/yr$ (corresponding to $\Sigma_{SFR}=1.7\ M_{\odot}\ yr^{-1}\ kpc^{-2}$). Given that clump B is classified as `intermediate' by the N-BPT, we do not measure a SFR from \ha, as it may be contaminated by AGN ionization. The spectrum of the clumps is characterized by an almost featureless continuum as expected from the cluster age estimated from the EW(\ha). One might possibly analyze the Balmer lines in absorption in order to derive a more reliable stellar age, but with such a strong infilling emission (EW(\ha)$>$300$\AA$) one would require a much higher SNR in the continuum than actually available in order to reliably measure the Balmer absorptions. Actually, the emission is so strong that completely obliterates any absorption in these clumps.
We will further discuss in the next Section why these particular locations are so special, and what has probably triggered the star formation in the clumps.

\section{Outflowing gas and positive feedback} \label{jetpf}
\
The presence of high velocity gas in the EELR region is revealed by the ionized gas dynamics traced by our MUSE observations. The nuclear spectrum shown in Fig.~\ref{spectra} presents an asymmetric profile of the [OIII] lines. The [OIII] lines are an ideal tracer of extended outflowing ionized gas, as they cannot be produced in the high-density, sub-parsec scales typical of AGN Broad-Line Regions (BLR), and have been widely used in the literature to study the winds properties (e.g. Villar-Mart\'in et al. \citealp{villar11}, Zhang et al. \citealp{zhang11}, Brusa et al. \citealp{brusa15}). A similar, asymmetric profile is observed for other forbidden lines, as the [NII] and [SII] doublets.

\begin{figure}
	\begin{center}
		\includegraphics[width=0.5\textwidth]{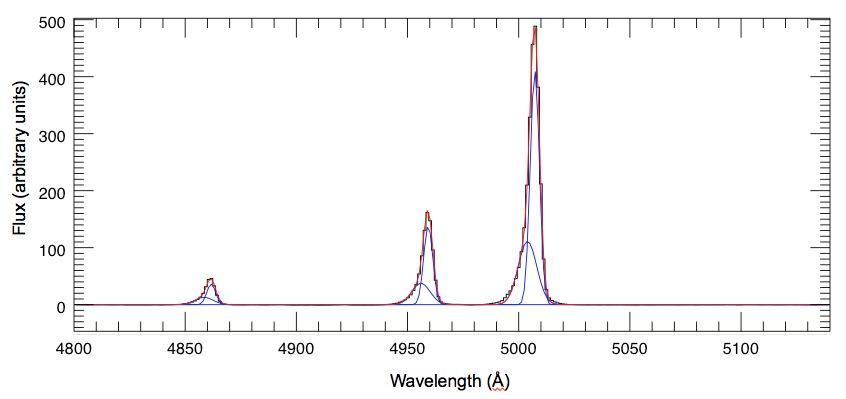}
	\end{center}
	\caption{Two components Gaussian fitting of the $1.6"\times1.6"$ rest frame nuclear spectrum of NGC~5643 for the \hb,\oiiia\ and \oiiib\ emission lines. The \oiiib\ lines in the nucleus shows an asymmetric blue wing, reproduced with blueshifted Gaussian components. The systemic components have a $FWHM_{sys}=286\pm 6\ km\ s^{-1}$, while the outflowing components have a centroid shifted by $v_{0,out}=-200\ km\ s^{-1}$ and a $FWHM_{out}=590\ km\ s^{-1}$. The projected velocity at the 10th percentile of the overall emission-line profile is $v_{10}=-430\ km\ s^{-1}$, confirming the presence of outflowing, high velocity gas from the nuclear region.}
	\label{o3fit}
\end{figure}

We fit the \hb, \oiiia\ and \oiiib\ nuclear spectrum of NGC~5643 (in a region of $1.6"\times1.6"$) with a double Gaussian for each line, a systemic component and an outflowing one (see Fig.~\ref{o3fit}). All the lines are fitted simultaneously, fixing the wavelengths ratios and [OIII] flux ratios according to theoretical values. The systemic components have a $FWHM_{sys}=286\pm 6\ km\ s^{-1}$, while the outflowing components have a centroid shifted by $v_{0,out}=-200\pm15\ km\ s^{-1}$ and a $FWHM_{out}=590\pm20\ km\ s^{-1}$. The projected velocity at the 10th percentile of the overall emission-line profile of the best fitting model is $v_{10}=-430\ km\ s^{-1}$. These high velocities confirm the presence of outflowing gas from the nuclear region.

\begin{figure}
	\begin{center}
		\includegraphics[width=0.5\textwidth]{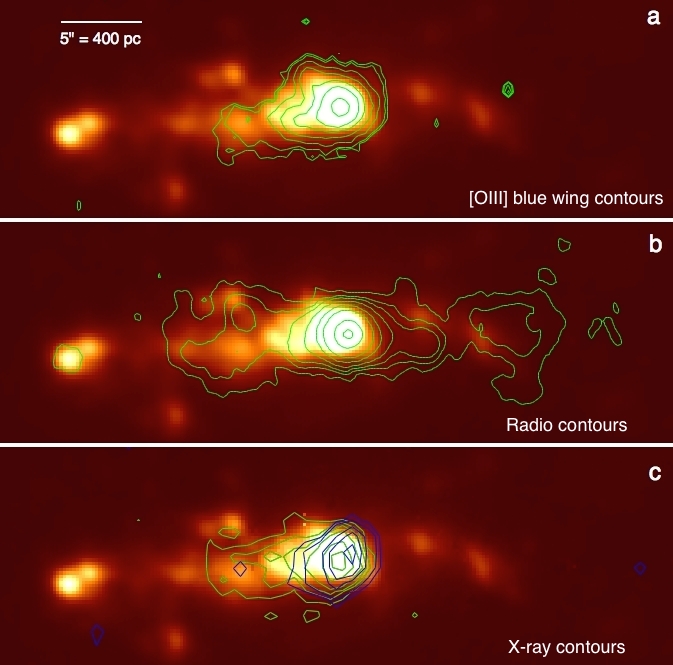}
	\end{center}
	\caption{The outflowing gas in the core of NGC~5643. \textit{Panel a:} \ha\ map of the central region, with superimposed the contours of the flux in the blue wing ($v<-400\ km\ s^{-1}$) of the \oiiia\ emission line, tracing the high velocity outflowing ionized gas. \textit{Panel b:} \ha\ map with superimposed the 8.4 GHz radio contours from the VLA observations presented in Leipski et al. \citealp{leipski06}. The diffuse radio jet is coincident with the location of the blue \oiiia\ wing. \textit{Panel c:} \ha\ map with superimposed Chandra X-ray contours, 0.3-0.5 keV shown in green, and 0.5-8 keV shown in blue. While the high energy data are located in the nuclear region only, the lower energy channels show extended emission compatible with an outflowing jet. Again, the location of the jet traced by X-rays corresponds to the ionized gas and radio emitting gas, and points towards the star forming blue clump, suggesting a possible connection between the star forming episode and the outflowing material.}
	\label{jet}
\end{figure}

Thanks to our IFU data, we are able to spatially map this high velocity gas: the upper panel of Fig.~\ref{jet} shows the contours of high velocity \oiiib\ emitting gas, derived by integrating the \oiiib\ emission in the spectral channels with $v<-400\ km\ s^{-1}$ with respect to the bulk velocity of the systemic component. The contours are superimposed to the \ha\ map. \\

NGC~5643 is also known to host a double sided diffuse radio jet, aligned with the bar, which extends by $\sim 15"=1.2\ kpc$ in both the E and the W direction. The radio jet is observed by Morris et al. \cite{morris85} at 6 and 20 cm, and by Leipski et al. \cite{leipski06} at 3.5 cm. The higher resolution ($1.7"\times1.7"$ beam) 3.5 cm VLA contours are shown in Fig.~\ref{jet}, central panel, superimposed to the \ha\ emission line map from our MUSE data.
The radio emission morphology is similar to the EELR, 
although the \ha\ emission is suppressed in the W direction due to the higher extinction. 
Given the presence of a low luminosity radio jet, the high velocity ionized gas could be accelerated by a sort of ``snowplough'' effect, with the high velocity shocks induced by the outflowing plasma heating and compressing the external gas (see e.g. Axon et al. \citealp{axon98}, Capetti et al. \citealp{capetti99}).

An alternative mechanism to explain the detected radio emission could be the interaction between an accretion disk wind and the surrounding interstellar medium (see e.g. Faucher-Gigu\`ere \& Quataert \citealp{fg12}, Zakamska \& Greene \citealp{zakamska14}, Nims et al. \citealp{nims15}, Harrison et al. \citealp{harrison15}).  We note in fact that the radio luminosity of NGC~5643 is $\nu L_{\nu}[8.4\ GHz]=0.55 \cdot 10^{21}\ erg\ s^{-1}$ (Leipski et al. \citealp{leipski06}), requiring a kinetic energy of the outflow of $L_{wind} \sim 2\cdot10^{39}\ erg\ s^{-1}$ if the efficiency of conversion in AGN driven winds is comparable to that in supernova driven winds. Thus the kinetic luminosity of the wind would be $\sim1\%$ of the bolometric luminosity oif NGC~5643, making this scenario viable. In any case, the radio emission is consistent with being produced by a low luminosity radio jet or quasar wind that are interacting with the ISM in the host galaxy. \\

We also show the X-ray images obtained filtering the Chandra data in the 0.3-0.5 keV and 0.5-8 keV energy bands in Fig.~\ref{jet}, bottom panel, where the high energy data are shown in blue and the lower energy data are shown in green. 
The high energy X-ray emission is concentrated in the central region, mostly in the unresolved central point source, with a minor fraction spread over a few arcsec around the centre. This is analogous to what has been observed in other similar sources (e.g. NGC~1365, Wang et al. \citealp{wang09}; NGC~4151, Wang et al. \citealp{wang10}). The unresolved emission is due to the direct emission of the central AGN (including the primary emission and the reflected component by the circumnuclear gas on parsec scale). The spatially resolved hard emission is expected to be due to a combination of reflection of the primary AGN emission by farther circumnuclear gas, at a distance of a few hundred pc, and the contribution of galactic X-ray binaries. Disentangling these components would require a higher S/N observation than available here, and will become possible with a forthcoming long {\em Chandra} observation scheduled in 2015. 
The soft emission is instead dominated by the extended, spatially resolved component, and is due to the emission of hot gas (typical temperatures kT$\sim$0.2-1~keV) either shock-heated, or photoionized by the central AGN. In particular, the spatial coincidence of the soft X-ray emission and the [OIII] emission points to a common origin of the two components, both due to photoionization by the central AGN of a two-phase gas, with the denser one associated to the [OIII] emission and the less dense, more photoionized one responsible for the X-ray emission  (e.g. Bianchi et al. \citealp{bianchi06}, Balmaverde et al. \citealp{balma12}).\\

The outflow direction as detected by radio, X-ray and ionized gas dynamics is pointing towards the location of the blue, \ha\ bright clumps, suggesting a mutual connection between the star formation in the clumps and the outflowing gas. In fact, we have seen in Sect.~\ref{lines} that the star forming clumps A and B are not only on the path of the material outflowing from the nucleus, but they are also located at the receding edge of the dust lane in the bar. Bars are in fact preceded by dust lanes in their motion, which have been recognized as related to shocks in the gas flow, and correspond to the regions of highest gas density  (e.g. Athanassoula \citealp{athana92}). Despite the high density, the gas flow is accompanied by very strong shear that inhibit star formation. Such shocks are usually found in the leading edge of the bar, roughly parallel to its major axis. \textrm{H~II} regions and star formation are instead preferentially offset from the dust lane, towards the leading side, as well as at the end of the bar major axis, where secondary gas density enhancements are seen both in simulations and observationally (e.g. Sheth et al. \citealp{sheth02}). As clumps A and B are located on the trailing edge of the dust lane, they are not at the expected position for \textrm{H~II} regions surrounding a bar, reinforcing the hypothesis of a mutual connection with the nuclear outflow. In this scenario, the outflow has been compressing the gas in the dust lane at the intersection point, triggering star formation (``positive feedback''). 


\section{Conclusions}

\begin{figure}
	\begin{center}
		\includegraphics[width=0.5\textwidth]{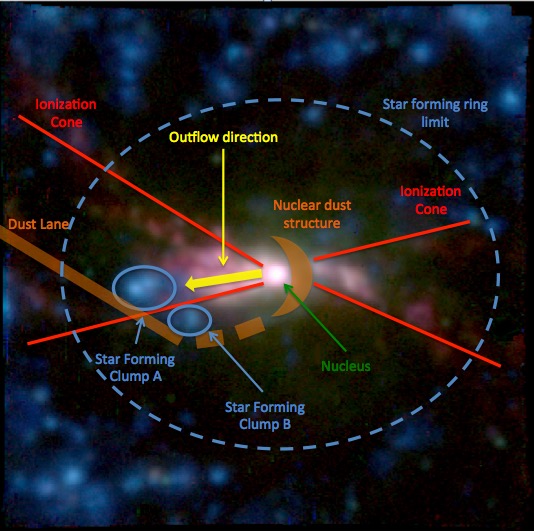}
	\end{center}
	\caption{Schematic view of the main structures revealed by MUSE in the central region of NGC~5643, drawn over the three colors image from Fig.~\ref{3colors}, panel c. The ionization cone borders are highlighted with red lines, the dust lane and nuclear dust structure location in brown. The two star forming clumps A and B are marked with blue circles, and the dashed blue ellipse, corresponding to a circle with the same inclination of the galaxy, shows the inner radius of the star forming ring around the nuclear region of the galaxy.}
	\label{schema}
\end{figure}
We have presented MUSE integral field data of NGC~5643, a local barred Seyfert 2 galaxy at a distance of $\sim17.3\ Mpc$. The large wavelength coverage allows us to study the main optical emission lines in the spectra, and map the ionization and dynamics of the gas. 

We detected a double sided ionization cone due to AGN radiation, almost parallel to the galaxy bar. The cone may be collimated by a dusty structure of $580\times125\ pc$ around the nucleus, connected with the dust lane in the bar and dominated by shock ionization. At the center of the cone an outflow is revealed by the high velocity, [OIII] emitting gas, as well as in radio and X-ray data. The ionized gas is moving away from the nucleus at projected velocities up to $v_{10}=-432\ km/s$. 

Two \ha\ bright, star formation dominated clumps, located at the receding edge of the dust lane in the bar, where the gas density is highest, are located at the location where the fast outflowing gas from the central AGN encounter the dense material on the bar, strongly suggesting a relation between the two phenomena. The clumps are the youngest in the MUSE field of view, as estimated from their \ha(EW), and are the only star formation dominated regions closer to the nucleus ($\sim1.2\ kpc$) than the star forming ring located at the edge of the field of view ($\sim 2.3\ kpc$, see Fig.~\ref{schema}). We propose that the star formation in the clumps (SFR=$0.03\ M_{\odot}/yr$ for clump A) is due to `positive feedback' of the AGN outflow, that is compressing the gas at the dust lane edge inducing star formation (see Fig.~\ref{schema}).

The presented data suggest that positive feedback may be present also at low AGN luminosity and outflow energetic. The upcoming observations in the framework of the MAGNUM survey will allow to assess if it represents an important ingredient in the complex black hole-host galaxy coevolution also in the local Universe.

\section*{Acknowledgments}

MUSE data were obtained from observations made with the ESO Telescopes at the Paranal Observatories. We are grateful to the ESO staff for their work and support. We are grateful to C. Leipski for providing us the VLA radio data of NGC~5643. GC, AM and SZ acknowledge support from grant PRIN-INAF 2011 ``Black hole growth and AGN feedback through the cosmic time'' and from grant PRIN-MIUR 2010-2011 ``The dark Universe and the cosmic evolution of baryons: from current surveys to Euclid''. SZ and AG have been supported by the EU Marie Curie Integration Grant ``SteMaGE'' Nr. PCIG12-GA-2012-326466  (Call Identifier: FP7-PEOPLE-2012 CIG). AG acknowledges support from the European Union FP7/2007-2013 under grant agreement n. 267251 (AstroFIt). MB acknowledges support from the FP7 Career Integration Grant ``eEASy'' (CIG 321913).

\label{lastpage}

\end{document}